\documentclass{rcdj}
\begin{document}

\setcounter{page}{29}
\journal{REGULAR AND CHAOTIC DYNAMICS, V.\,9, \No1, 2004}
\runningtitle{NOTES ON DIFFUSION IN COLLISIONLESS MEDIUM}
\title{NOTES ON DIFFUSION IN COLLISIONLESS MEDIUM}
\runningauthor{V.\,V.\,KOZLOV}
\authors{V.\,V.\,KOZLOV}
{Steklov Institute of Mathematics\\
Russian Academy of Sciences\\
117966 Moscow, Russia\\
E-mail: kozlov@pran.ru}

\abstract{A collisionless continuous medium in Euclidean space is discussed,
i.\,e. a continuum of free particles moving inertially, without
interacting with each other. It is shown that the distribution density of
such medium is weakly converging to zero as time increases indefinitely.
In the case of Maxwell's velocity distribution of particles, this density
satisfies the well-known diffusion equation, the diffusion coefficient
increasing linearly with time.}
\amsmsc{37H10, 70F45}
\doi{10.1070/RD2004v009n01ABEH000262}
\received 06.10.2003.

\maketitle

\section{Dynamics of collisionless medium in the Euclidean space}

We are going to consider a very simple object~--- \textit{a collisionless
continuous medium}, i.\,e. a continuum of free particles moving
inertially, without interacting with each other. The configurational space
of a particle is the~$n$-dimensional Euclidean space~$\mR^n$ with
orthogonal coordinates~$x_1,\ldots,x_n$;
let~$\mR^n=\{\omega_1,\ldots,\omega_n\}$ be the velocity space. The direct
product~$\mR^n_x\times\mR^n_\omega=\Gamma$ is the phase space of a free
particle.

Let~$\rho(\omega,x)$ be the particle distribution density at the initial
time~$t=0$. We can assume that the density is normalized to the total mass
of the collection of particles (the total  mass is supposed to be finite). In
other words,~$\rho$ is the density of some probability
measure$\colon$~$\rho\geqslant0$ and
$$
\intl_\Gamma\rho{d}^n\omega{d}^nx=1.
$$
 From the very beginning, it is possible to use the probabilistic approach
and treat a collisionless medium as a \textit{Gibbs ensemble} of
identical systems, where each system is a free particle in the Euclidean
space~$\mR^n$.

According to the elementary principles of statistical mechanics,
density~$\rho_t$ at time~$t$ is given by
$$
\rho_t(\omega,x)=\rho(\omega,x-\omega{t}).
$$
It is clear that~$\rho_0=\rho$. It is also clear
that
$$
u(x,t)=\intl_{\mR^n}\rho(\omega,x-\omega{t})d^n\omega\eqno{(1.1)}
$$
is the density of the collisionless medium at point~$x$ at time~$t$. Our
objective is to study the concentration~$u$, its evolution and limit behavior
as~$t\rightarrow\pm\infty$.

In Ref.~[1], the problem of  evolution of a collisionless
medium inside a box with mirror walls was discussed. Upon a  simple regularization,
this problem can be reduced to the problem with periodic boundary
conditions$\colon$ the configurational space~$\mR^n=\{x\}$ is factorized using the lattice~$(2\pi\mZ)^n$. As a result, the phase space~$\Gamma$ is the
direct product of torus~$\mT^n=\{x_1,\ldots,x_n{mod}2\pi\}$
and~$\mR=\{\omega\}$.
\begin{teo}
Let~$\rho\in{L}_1(\Gamma)$ and $\varphi$ be the characteristic function of
a bounded measurable region~$D\subset\mR^n=\{x\}$. Then
$$
\intl_{\mR^n}u(x,t)\varphi(x)d^nx=\intl_Du(x,t)d^nx\rightarrow{0}\eqno{(1.2)}
$$
as $t\rightarrow\pm\infty$.
\end{teo}

This theorem is intuitively obvious$\colon$ the particles scatter to
infinity, each with its own velocity, and, therefore, their concentration
in any finite region of~$\mR^n=\{x\}$ is decreasing indefinitely.

In fact, (1.2) holds for any essentially bounded measurable
function~$\varphi:\mR^n\rightarrow\mR$. Equation~(1.2) has the following
meaning$\colon$ weak density limit~$\rho_t$ is zero
as~$t\rightarrow\pm\infty$. This assertion was  proven in Ref.~[1]
for a compact configurational space. In any case,
$$
\lim_{t\rightarrow\pm\infty}\intl_\Gamma\rho(\omega,x-\omega{t})\varphi(x)d^nxd^n\omega=\int_\Gamma\bar{\rho}\varphi{d}^nxd^n\omega,\eqno{(1.3)}
$$
where $\bar{\rho}$ is the Birkhoff average of~$\rho$$\colon$
$$
\bar{\rho}(\omega,x)=\lim_{\tau\rightarrow\infty}\frac{1}{2\tau}\intl^\tau_{-\tau}\rho(\omega,x-\omega{t})dt.
$$
It is easy to calculate$\colon$
$$
\bar{\rho}=\lim_{\tau\rightarrow\infty}\frac{1}{2\omega\tau}\intl^{x+\omega\tau}_{x-\omega\tau}
\rho(\omega,\xi)d\xi=0
$$
for almost all~$\omega$, because (according to Fubini's theorem) the
integral of~$\rho(\omega,\xi)$ over the variable~$\xi\in\mR$ exists (and
is finite) for almost all~$\omega$. Specifically, the integral~(1.3) is
also zero.

The question whether~$u(x,t)$ itself tends to zero
as~$t\rightarrow\pm\infty$ is somewhat more interesting. We discuss it for
the case where~$\rho$ is the product of two summable functions~$h(\omega)$
and~$\varphi(x)$. Then, the integral~(1.1) takes the form
$$
u(x,t)=\intl_{\mR^n}h(\omega)\varphi(x-\omega{t})d^n\omega.\eqno{(1.4)}
$$

Let us present~$\varphi$ in terms of Fourier transformation
$$
\varphi(z)=\frac{1}{(\sqrt{2\pi})^n}\intl_{\mR^n}\Phi(\xi)e^{i(z,\xi)}d^n\xi,
$$
and put
$$
H(z)=\frac{1}{(\sqrt{2\pi})^n}\int_{\mR^n}h(\omega)e^{-i(z,\omega)}d^n\omega.
$$

\begin{teo}
Suppose that~$\Phi$ is a bounded summable function and
~$H\in{L}_1(\mR^n)$. Then, for all~$x\in\mR^n$, function~$(1.4)$ tends
to zero as~$t\rightarrow\pm\infty$.
\end{teo}

Indeed, according to Fubini's theorem,
$$
u=\frac{1}{(\sqrt{2\pi})^n}\intl_{\mR^n}\Phi(\xi)e^{i(x,\xi)}\left[\intl_{\mR^n}
h(\omega)e^{-it(\omega,\xi)}d^n\omega\right]d^n\xi=\intl_{\mR^n}\Phi(\xi)H(t\xi)e^{i(x,\xi)}d^n\xi.
$$
When $t>0$, this integral is equal to
$$
\frac{1}{t}\intl_{\mR^n}H(z)\Phi(\frac{z}{t})e^{i(x,z)/t}d^nz,
$$
which is of order~$O(1/t)$  as~$t\rightarrow\infty$ (by the assumption of the theorem).

\section{Heat conduction equation}

The equation is
$$
u_\tau=\sigma^2\triangle{u},\eqno{(2.1)}
$$
where $\tau$ is a time variable, $\triangle$ is the Laplace operator,
while~$\sigma=\const$. Equation~(2.1) is a special form of the
diffusion equation; $\sigma^2$ is the diffusion coefficient.

The solution of~(2.1) is well-known
$$
u(x,\tau)=\frac{1}{(2\sigma\sqrt{\pi\tau})^n}\intl_{\mR^n}e^{\frac{|x-\xi|^2}
{4\sigma^2\tau}}
\varphi(\xi)d^n\xi,\,\tau>0,
\eqno{(2.2)}
$$
where $\varphi$ is the initial temperature distribution,
and~$|q|=q^2_1+\ldots+q^2_n$. The function~$\varphi$ is customarily supposed to
be \textit{continuous} and \textit{bounded\/}. The latter condition
ensures convergence of integral~(2.2), while the continuity property
allows proving that
$$
\lim_{\tau\rightarrow0}u(x,z)=\varphi(x).
$$
The point is that the exponential term in (2.2) (together with the
term outside the integral) tends to the delta-function~$\delta(x-\xi)$
as~$\tau\rightarrow0$.

However, the integral~(2.2) also converges on the assumption of
summability of~$\varphi$ (i.\,e. when ${\varphi\in{L}_1)}$.
It turns out that~(2.2) can be presented in the form of~(1.4), and this, among other
things, implies that~$u(x,0)=\varphi(x)$ if~$\varphi$ is
summable.

Let $\tau=t^2/2$ and $\xi=x-\omega\tau$,
where~$\omega=(\omega_1,\ldots,\omega_n)$. Then,~$d\xi_j=-td\omega_j$ and
$$
\begin{aligned}
&\frac{1}{(\sigma\sqrt{2\pi}t)^n}\intl_{\mR^n}e^{-\frac{|x-\xi|}{2\sigma^2t}}\varphi(\xi)d^n\xi=\\
&=\frac{(-1)^n}{(\sigma\sqrt{2\pi})^n}\intl^{-\infty}_\infty\ldots\intl^{-\infty}_\infty{e}^{-\frac{|\omega|^2}{2\sigma^2}}\varphi(x-\omega{t})d^n\omega=\\
&=\frac{1}{(\sqrt{2\pi}\sigma)^n}\intl_{\mR^n}{e}^{-\frac{|\omega|^2}{2\sigma^2}}\varphi(x-\omega{t})d^n\omega.
\end{aligned}\eqno{(2.3)}
$$
In particular, $u(x,0)=\varphi(x)$.

Thus, if we adopt the normal law of velocity distribution
$$
h(\omega)=\frac{1}{(\sqrt{2\pi}\sigma)^n}{e}^{-\frac{|\omega|^2}{2\sigma^2}},\eqno{(2.4)}
$$
then the density~$u(x,t)$ of the collisionless medium, given by the
integral~(1.4), satisfies the diffusion equation
$$
u_t=t\sigma^2\triangle{u}.\eqno{(2.5)}
$$
The diffusion coefficient~$t\sigma^2$ increases indefinitely with time. As
distinct from the heat conduction equation, this equation is invariant
under time reversion~$t\mapsto-t$. This reflects the
property of reversibility of the  equations of motion for a free particle. Specifically,
concentration of particles at any point~$x\in\mR^n$ decreases indefinitely
both as~$t\rightarrow +\infty$ and as~$t\rightarrow-\infty$.

From the statistical mechanics point of view, it would be more appropriate
to treat the distribution~(2.4) as a Maxwell distribution,
dispersion~$\sigma^2$ being proportional to the absolute temperature of
the gas. This distribution does not vary with time (because the medium is
collisionless), and the temperature field is proportional to the density
of the collisionless medium (after identifying~$t^2/2$ with~$\tau$).

The simple equation~(2.3) is also useful for the analysis of heat
propagation in~$\mR^n$. For example, let~$\vfi>0$ inside an open bounded
region~$D\subset\mR^n$ and~$\vfi=0$ outside this region. Then~$u(x,t)>0$
at any  point~$x\in\mR$ for arbitrarily small~$t>0$. This property immediately follows from the law of motion of a collisionless medium $\colon$ however distant a point~$x\in\mR^n$ may be, it will be reached
in an arbitrarily small time by very fast particles, located initially
in~$D$.

\section{An example of a nonstandard diffusion equation}

It would be a mistake to think that functions of the form~(1.4) satisfy the diffusion equation in its well-known form. Let us put, for
example,~$n=1$ and
$$
h(\om)= e^\om\,\text{ when}\,\,\om\leqslant0\quad\text{and}\quad
h(\om)=0\,\text{ when}\,\,\om>0.
$$
Then~(1.4) becomes
$$
u(x,t)= \intl_{-\infty}^0e^\om\vfi(x-\om t)\,d\om.\eqno{(3.1)}
$$
Integrating by parts yields the equation
$$
u=tu_x+\vfi(x),\eqno{(3.2)}
$$
which does not contain the derivative of~$u_t$ at all. Putting~$t=0$
in~(3.2), we find that~$u(x,0)=\vfi(x)$.

However, Theorem~2 cannot be applied
straightforwardly because
$$
H(z)=\frac{1} {\sqrt{2\pi}}\intl_{-\infty}^0e^\om
e^{-iz\om}\,d\om=\frac{1}{\sqrt{2\pi} (1-iz)}
$$
does not belong to~$L_1(\mR)$. However, if, instead of the boundedness
of~$|\Phi(z)|$, we require that this function vanishes at infinity
as~$O(|z|^{-\al}),\al>0$, then again we can say that the integral~(3.1)
tends to zero as~$t\rightarrow\pm\infty$, for each~$x\in\mR$.

The function~$\vfi$ can be eliminated from~(3.2) if we replace~(3.2) by
the equation
$$
u_t= (tu_x)_t\eqno{(3.3)}
$$
and~the Cauchy condition $u|_{t=0}=\vfi(x)$. Thus,~(3.3) should also be
considered  a diffusion equation.

Equation~(3.3) is not invariant under time reversal. This fact can be
easily explained$\colon$ all the particles of a collisionless medium move
to the left. To have symmetry between the past and the future, one should
assume that there is symmetry between "left" and "right" in the distribution of velocities. We come to a simple
\begin{pro*}
If $h(-\om)=h(\om)$, then $u(x,-t)=u(x,t)$.
\end{pro*}
Indeed,
\begin{gather*}
u(x,-t)=\intl_{-\infty}^\infty h(\om)\vfi(x+\om t)\,d\om=-\intl_\infty
^{-\infty}h(-\om')\vfi(x-\om't)\,d\om'=\\
=\intl_{-\infty}^\infty h(\om)\vfi(x-\om t)\,dt=u(x,t).
\end{gather*}

Let, for example,
$$
h(\om)=\frac{1}{2}e^{-|\om|}.
$$

Integrating by parts, we obtain
\begin{gather*}
\intl_{-\infty}^0e^\om\vfi(x-\om t)\,d\om=\vfi(x)+t\vfi'(x)+t^2
\intl_{-\infty}^0e^\om\vfi''(x-\om t)\,d\om,\\
\intl_0^\infty e^{-\om}\vfi(x-\om t)\,d\om=\vfi(x)-t\vfi'(x)+t^2
\intl_0^\infty e^{-\om}\vfi''(x-\om t)\,d\om.
\end{gather*}
This leads to
$$
u=\vfi(x)+\frac{t^2}{2}u_{xx},
$$
with $u|_{t=0}=\vfi(x)$. This equation is equivalent to the equation of
evolution
$$
u_t=\left(\frac{t^2}{2}u_{xx}\right)_t,
$$
with the initial condition~$u|_{t=0}=\vfi(x)$. Unlike~(3.3), this equation  is
invariant under time reversal.

\section{Diffusion in the compact case}

Now, let the configurational space be an~$n$-dimensional
torus~$\mT^n=\{x_1,\ldots,x_n\mod2\pi\}$. In this case, the density of a
collisionless medium is also given by~(1.1). As it was shown in~[1], the
weak limit of~$u(x,t)$ as~$t\rightarrow\pm\infty$ is equal to
$$
\overline{u}=\frac{1}{(2\pi)^n}\intl_{\mT^n}\intl_{\mR^n}\rho(\om,x)\,
d^n\om\,d^nx.\eqno{(4.1)}
$$

Under certain additional conditions, we can state that~$u(x,t) \rightarrow
\overline{u}$ as~$t\rightarrow\pm\infty$ for all~$x\in\mR^n$. To this end,
suppose that the density~$u(x,t)$ is given by the
integral~(1.4). Let
$$
\sum\vfi_me^{i(m,x)},\quad m\in\mZ^n \eqno{(4.2)}
$$
be the Fourier series of a bounded measurable function~$\vfi\colon$
$\mT^n\rightarrow\mR$.
\begin{teo}
Suppose that the series
$$
\sum|\vfi_m|\eqno{(4.3)}
$$
converges; then~$u(x,t)\rightarrow\overline{u}$ as~$t\rightarrow\pm\infty$
for all~$x\in\mT^n$.
\end{teo}
Inserting~(4.2) into~(1.4), we obtain$\colon$
$$u(x,t)=\overline{u}+\suml_{m\neq0}\vfi_me^{i(m,x)}\intl_{\mR^n}h(\om)e^
{-i(m,\om)t}\,d^n\om,\eqno{(4.4)}
$$
where
$$
\overline{u}=\vfi_0\intl_{\mR^n} h(\om)\,d^n\om
$$
coincides with~(4.1). Because of the convergence of the series~(4.3)  the summation over~$m$ and integration
over~$\om$ can be interchanged.

Let~$\eps$ be an arbitrary small positive number. Since the series~(4.3)
converges, there exists a number~$N$, depending on~$\eps$, such that
\begin{gather*}
\left|\suml_{|m|>N}\vfi_me^{i(m,x)}\intl_{\mR^n}h(\om)e^{-i(m,\om)t}\,
d^n\om\right|\leqslant\\
\leqslant\left(\suml_{|m|>N}|\vfi_m|\right)\intl_{\mR^n}h(\om)\,d^n\om
\end{gather*}
is less than~$\eps/2$. Then, the
terms with indices subject to  $|m|\leqslant N$ and~$m\neq0$ tend to zero
as~$t\rightarrow\infty$, because (by the Riemann--Lebesgue theorem) so does the integral
$$
\intl_{\mR^n}h(\om)e^{-i(m,\om)t}\,d^n\om
$$
 Thus, there exists~$T(\eps)$ such that for ~$t>T$ the sum of this finite number of terms is
less than~$\eps/2$, which is the required result.

The work was supported by the Russian Foundation for Basic Research
(grant~02-01-01059) and the Foundation for Leading Scientific Schools
(grant~136.2003.1).


\begin{thebibliography}{1}
\bibitem{k1}
    \author{V.\,V.\,Kozlov}
    \title{Kinetics of collisionless continuous medium}
    \journal*{Reg. \& Chaot. Dyn.}
    \year{2001}
    \volume{6}
    \no{3}
    \page{235--251}
    \\
        \mbox{}\\
            \mbox{}


\end{thebibliography}
\end{document}